\documentclass[english,prb,twocolumn,nofootinbib]{revtex4}

\usepackage{changes}
\usepackage{cancel}

\usepackage{amssymb,amsmath,graphicx}
\usepackage{bm}
\usepackage{hyperref}
\usepackage[normalem]{ulem}

\hypersetup{colorlinks=true,linkcolor=blue,citecolor=violet,urlcolor=blue}
\usepackage{tikz}
\usetikzlibrary{decorations.pathmorphing,decorations.markings,trees,positioning,arrows}
\usetikzlibrary{calc,patterns}
\usetikzlibrary{shapes,backgrounds}

\newcommand{\non}{\nonumber\\}

\renewcommand{\varpi}{\epsilon}
\newcommand{\sign}{\text{sign}}

\begin{document}
	\title{Quantum critical Eliashberg theory, the SYK superconductor and their holographic duals}
	
	\author{Gian-Andrea Inkof}
	
	\affiliation{Institute for Theory of Condensed Matter, Karlsruhe Institute of Technology 76131 Karlsruhe, Germany}
	\author{Koenraad Schalm}
	\affiliation{
		Instituut-Lorentz, $\Delta$-ITP, Universiteit Leiden, P.O. Box 9506, 2300 RA Leiden, The Netherlands
	}
	\author{J\"{o}rg Schmalian}
	\affiliation{Institute for Theory of Condensed Matter, Karlsruhe Institute of Technology 76131 Karlsruhe, Germany}
	\affiliation{Institute for Quantum Materials and Technologies,
		Karlsruhe Institute of Technology, Karlsruhe 76021, Germany}
	
	\begin{abstract}
		Superconductivity is abundant near quantum-critical points, where fluctuations suppress the formation of Fermi liquid quasiparticles and the BCS theory no longer applies. Two very distinct approaches have been developed to address this issue: quantum-critical Eliashberg theory and holographic superconductivity. The former includes a strongly retarded pairing interaction of ill-defined fermions, the latter is rooted in the duality of quantum field theory and gravity theory. We demonstrate that both are different perspectives of the same theory. We derive holographic superconductivity in form of a gravity theory with emergent space-time from a quantum many-body Hamiltonian - the Yukawa SYK model - where the Eliashberg formalism is exact. Exploiting the power of holography, we then determine the dynamic pairing susceptibility of the model. Our holographic map comes with the potential to use quantum gravity corrections to go beyond the Eliashberg regime.
	\end{abstract}
	\maketitle
	
	\section{Introduction}
	Superconductivity is the natural ground state of a clean metal with Fermi liquid properties. A cornerstone of paired-electron superconductivity in Fermi liquids is Eliashberg theory \cite{Eliashberg1960,Eliashberg1961}, which extends BCS theory by accounting for the dynamics of soft boson quanta that mediate the electron-electron interaction.  
	Experimentally, superconductivity is particularly abundant near quantum critical points\cite{Mathur1998,Kasahara2010,Sachdev2011}; yet here electrons no longer form a Fermi liquid and the Cooper instability paradigm does not apply. Indeed, repeating  Cooper's analysis for a non-Fermi liquid but with instantaneous pairing interaction\cite{Balatsky1993,Sudbo1995} suggests, at first glance, that superconductivity should rather be the exception than the
	rule. Important progress was made by realizing that dynamical retardation
	effects - naturally present in  the Eliashberg formalism - are crucial in critical systems\cite{Abanov2001,Abanov2001b,Chubukov2005,She2009,Abanov2020,Wu2020,She2011}.  Examples  are gauge-field induced composite fermion pairing\cite{Bonesteel1996,Metlitski2015},
	color magnetic interaction in high-density quark matter\cite{Son1999,Chubukov2005},
	superconductivity due to magnetic\cite{Abanov2001,Abanov2001b,Roussev2001}
	or Ising nematic\cite{Metlitski2015,Raghu2015} quantum critical fluctuations,
	or pairing in $U(1)$ and $Z_{2}$ spin-liquid states\cite{Metlitski2015}. Strictly put, the justification of the quantum-critical Eliashberg theory is not always clear, but in a coarser sense it should  capture the most relevant dynamics and the scaling properties \cite{Chowdhury2019b,Chubukov2020}.

	\begin{figure*}[t]
		\centering
		\textbf{~}\includegraphics[width=0.7\textwidth]{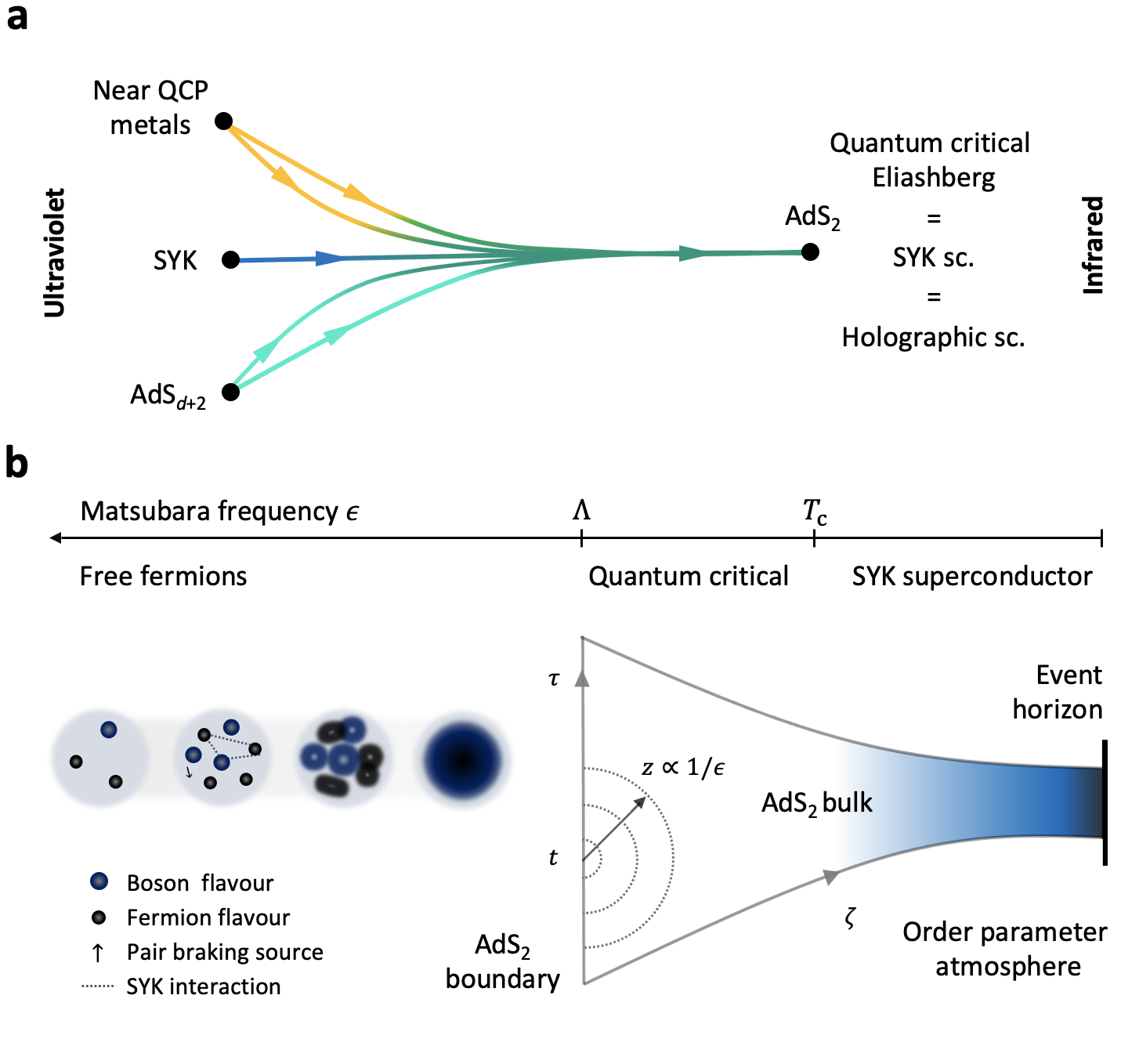}
		\caption{\textbf{The emergence of the dual holographic theory from the SYK superconductor.}
			\textbf{Top a:} The semi-universal low energy holographic action equal to quantum critical Eliashberg theory describing the IR of holographic superconductors (AdS$_{d+2}$), SYK superconductors, and metals near quantum critical points.  
			\textbf{Bottom b: }Below the UV scale $\Lambda$, a strongly coupled quantum-critical fluid forms out of interacting fermions and bosons.  Below this scale the fluctuating order parameter becomes  a scalar field with emergent AdS$_2$ gravity and forms an atmosphere around a black hole event horizon. The frequency $\varpi$ of the relative time $\tau_1-\tau_2$ and the absolute time $t=(\tau_1+\tau_2)/2$ of SYK pairing fields $F\left(\tau_1,\tau_2\right)=\frac{1}{N}\sum_{i=1}^{N}c_{i\uparrow}\left(\tau_1\right)c_{i\downarrow}\left(\tau_2\right)$ determine the set of geodesic half circles that span the anti-de Sitter space $(\tau,\zeta)$ on the gravity side.}
		\label{fig:sketch}
	\end{figure*}
	
	A tremendous amount of new insight into	quantum critical systems has been gained from a fundamentally different perspective given by the holographic dual formulation in terms of a higher-dimensional Anti-de Sitter (AdS) space\cite{Zaanen2015,Ammon2015,Hartnoll2018}. This also extends to superconductivity. The holographic superconductor \cite{Hartnoll:2008vx,Gubser:2008px} can consistently describe order-parameter condensation in strongly coupled theories without quasiparticles and naturally encapsulates quantum critical points (QCPs) in addition to thermal transitions. Also here, these results are often pushed beyond the matrix large-$N$ theories for which holography is formally justified.
	Again, this should capture the most relevant dynamics, even though  reliable quantitative information is in essence limited to systems where additional symmetries constrain the holographic model enough, e.g. 
	the canonical example of $\mathcal{N} = 4$ super Yang-Mills \cite{Maldacena:1997re}, or, as of relevance here, the (0+1)-dimensional model of Sachdev-Ye-Kitaev (SYK)~\cite {Sachdev1993, Sachdev2010, Kitaev2015}.

	In this article we show how the mutually complementary regimes of quantum critical Eliashberg theory and holographic superconductivity actually overlap and 	are, in fact, different perspectives of the same identical physics.	We derive holographic superconductivity in form of a gravity  theory with emergent  space-time from a quantum many-body Hamiltonian where the Eliashberg formalism is exact.
	To this end we focus on the quantum-critical transition to superconductivity in SYK-like models.
	
	In the normal state, the SYK model is described by a low-energy effective action, the so-called Schwarzian theory, that can also be derived from two-dimensional gravity with a scalar field (the dilaton)  \cite{Maldacena:2016hyu,Maldacena:2016upp}, where the same pattern of breaking of the conformal symmetry occurs. In fact, the structure of the Schwarzian action can be guessed with  considerations of conformal symmetry alone. In stark contrast, our quantitative map between SYK-Eliashberg theory and holography clearly cannot and does not rely on conformal symmetry: the presence of an explicit scale, the critical temperature $T_c$, breaks scale-invariance, and the dynamical equation of order parameter formation is sensitive to certain details of the system. Instead, the SYK-Eliashberg dynamics and its exact holographic equivalent describe the semi-universal RG-flow; semi-universal in the sense that it is grounded in a universal quantum critical SYK superconducting transition in the infrared (IR), but may have different microscopic, ultraviolet (UV) origins. This is sketched in Fig.~\ref{fig:sketch}.
	
	Using this exact map as an anchor, we exploit the power of holography to investigate the previously unknown dynamical pairing response of the SYK superconductor\footnote{This can be seen as an extension of the mapping of DC transport in Wilson-Fischer-type QCPs to holography, and using this map to predict finite-frequency responses near criticality \cite{Myers:2016wsu}.}.
	
	\section{From Eliashberg to holography}
	
	\subsection{\bf SYK-Eliashberg} 
	The explicit example of Eliashberg theory for a  tractable strongly coupled quantum critical point we shall focus on is the SYK superconductor constructed in ref.\cite{Esterlis2019} (see also \cite{Patel2018,Wang2019,Chowdhury2019,Hauck2020}). This is a generalization of the complex SYK model with $N$ charge-$q$, spin-1/2 fermions $c_{i\sigma}$ and four-fermion interactions to a more generic Yukawa interaction with $M$ soft bosons $\phi_{k}$
	\begin{eqnarray}
	H&=&-\mu\sum_{i=1,\sigma=\uparrow\downarrow}^{N}c_{i\sigma}^{\dagger}c_{i\sigma}+\frac{1}{2}\sum_{k=1}^{M}\left(\pi_{k}^{2}+\omega_{0}^{2}\phi_{k}^{2}\right)\nonumber \\
	&+&\frac{1}{N}\sum_{ijk\sigma}^{M,N}\left(g_{ij,k}+g^*_{ji,k}\right)c_{i\sigma}^{\dagger}c_{j\sigma}\phi_{k}\, .
	\label{eq:Hamiltonian}
	\end{eqnarray}
	In essence it encodes fermions coupled to a large number of Einstein phonons at a common frequency $\omega_0$ in the strong coupling limit.

	Disorder averaging over the interactions $g_{ij,k}$ around a zero mean can induce superconductivity in states without quasi-particles if $g_{ij,k}=(g_k)_{ij}$ are sampled from the Gaussian Orthogonal Ensemble rather than the Gaussian Unitary Ensemble with variance $\bar{g}^2$. Interpolating between the two ensembles by introducing a pair-breaking parameter $0\leq \alpha\leq1$ further allows to continuously vary the superconducting transition temperature $T_{\rm c}$ and tune it to vanish at a quantum critical point\cite{Hauck2020}. 
	
	These results follow  after a rewriting of the  path-integral in terms of bilocal propagator fields $G\left(\tau_{1},\tau_{2}\right)= \frac{1}{N}\sum_{i=1}^{N}c_{i\sigma}^{\dagger}\left(\tau_{1}\right)c_{i\sigma}\left(\tau_{2}\right)$ and $F\left(\tau_{1},\tau_{2}\right)=\frac{1}{N}\sum_{i=1}^{N}c_{i\uparrow}\left(\tau_{1}\right)c_{i\downarrow}\left(\tau_{2}\right)$ that correspond at the large-$N$ saddle point to the normal and anomalous Green's function of the Nambu-Eliashberg formalism.  Both have conjugate fields $\Sigma\left(\tau_{1},\tau_{2}\right)$ and $\Phi\left(\tau_{1},\tau_{2}\right)$ that play the role of self energies. The (Euclidean) action of the analysis is rather lengthy and can be found in the appendix.  In the large $N,M$ limit with $M/N$ fixed, the equations of motion are recognized as strongly coupled Eliashberg equations\cite{Esterlis2019,Patel2018,Wang2019,Chowdhury2019,Hauck2020}.
	
	In the normal state $F(\tau_{1},\tau_{2})=\Phi(\tau_{1},\tau_{2})=0$, and the theory flows to a SYK critical state defined by the time-translation invariant scaling dynamics with fermion and boson propagators
	\begin{align}
	\label{norm-state-sols}
	G_{\rm n}(\tau_1-\tau_2) 
	&=
	b_g\frac{\tanh\left(\pi q{\cal E}\right)+\sign(\tau_1-\tau_2)}{|\tau_1-\tau_2|^{\frac{1+\gamma}{2}}},
	\non
	D_{\rm n} (\tau_1-\tau_2) &= 
	b_d \frac{1}{|\tau_1-\tau_2|^{1-\gamma}}.
	\end{align}
	The exponent $0<\gamma<1$ can be tuned by varying the ratio $M/N$ and the charge density $n$ via Eq.\eqref{exponent_gamma} of the appendix. In the SYK literature one frequently uses $\Delta=(1+\gamma)/4$ instead of $\gamma$.
	The spectral asymmetry parameter ${\cal E}$ can be directly related to the particle density $n=\langle c_{i \sigma}^\dagger c_{i \sigma} \rangle $ via a generalized Luttinger theorem\cite{Georges2001,Wang2020}, where ${\cal E}(n=1/2)=0$. It is also linked to the density dependence of the zero-point entropy\cite{Sachdev2015} $2\pi{\cal E}=\frac{\partial S_{0}}{\partial n}$.
	While the two coefficients $b_{g,d}$ depend on details of the UV behavior, the important combination $\mu_\gamma\equiv \bar{g}^{2}b_{d}b_{g}^{2}=\tfrac{\gamma}{4\pi} \cosh^{2}\left(\pi q{\cal E}\right)\tan\left(\pi\gamma/2\right)$ is dimensionless and universal.  Thus, the normal state of the Hamiltonian \eqref{eq:Hamiltonian} is a strongly coupled, quantum-critical  fluid made of interacting fermions and bosons.
	
	In the superconducting state in the vicinity of the phase transition $F$ and  $\Phi$ remain small. We may then expand the action to second order and integrate out $\Phi$. This yields the leading contributions of superconducting fluctuations to the SYK action:
	\begin{eqnarray}
	S^{\left({\rm sc}\right)}/N&=&\int_{\omega,\varpi}\frac{F^{\dagger}
		\left(\omega,\varpi\right)F\left(\omega,\varpi\right)}{\Pi_{\rm n}\left(\omega,\varpi\right)} \label{eq:SYKscinit} \\
	&-&\frac{\bar{g}^{2}\lambda_{p}}{2}\int_{\omega,\varpi,\epsilon'}F^{\dagger}\left(\omega,\varpi\right)D_{\rm n}\left(\varpi-\varpi'\right)F\left(\omega,\varpi'\right).\nonumber
	\end{eqnarray} 
	Here, we Fourier transformed $F\left(\tau_1,\tau_2\right)$ with regards
	to the relative time $\tau_1-\tau_2$, with frequency $\varpi$,
	and absolute time $t=\left(\tau_1+\tau_2\right)/2$, with frequency
	$\omega$, respectively.  $\int_{\omega}\cdots=T\sum_{n}\cdots$
	stands for the sum over Matsubara frequencies.
	$\lambda_{p}=\left(1-\alpha\right)M/N$ modifies the coupling constant
	in the pairing channel and depends on the pair breaking strength $\alpha$.
	The first term in Eq.\eqref{eq:SYKscinit} is determined by the particle-particle
	response function 
	$\Pi_{\rm n}\left(\omega,\varpi\right)=G_{\rm n}\left(\frac{\omega}{2}-\varpi\right)G_{\rm n}\left(\varpi+\frac{\omega}{2}\right)$ 
	of the critical normal state. The second term  contains the singular
	pairing interaction with boson propagator $D_{\rm n}\left(\varpi\right)$. 
	
	Eq.\eqref{eq:SYKscinit} is then the IR effective action for the dynamic order parameter $F(\omega)_{\varpi}=F(\omega, \varpi)$ parametrized by a microscopic energy $\varpi$. In the dynamics of BCS theory the $\varpi$-dependence is neglected.
	The Eliashberg dynamics makes clear that $\epsilon$ is the relative energy of the fermions, and therefore representative of the characteristic energy at which the system is probed. One can indeed think of it as an RG scale, as our mapping to a holographic model will make clear.
	
	\subsection{Holographic map}
	\begin{table}
		\begin{ruledtabular}
			\begin{tabular}{ll}
				\textbf{Field-theory side}
				& \textbf{Gravity side}
				\\
				frequency $\epsilon$ of time lag $\tau_1-\tau_2$
				& 
				holographic dimension $\zeta$
				\\
				absolute time $(\tau_1+\tau_2)/2$
				& 
				time (Euclidean) $\tau$
				\\
				anomalous propagator $F$
				& 
				order parameter field $\psi$
				\\
				fermion bubble $\Pi_{\rm n}(\omega,\epsilon)$
				& $\partial_\tau \psi$, mass contrib. $m_{(1)}^2>0$
				\\
				pairing interaction  $D_{\rm n}(\epsilon)$
				& $\partial_\zeta \psi$, mass contrib. $m_{(2)}^2<0$
				\\
				Cooper pair charge $2q$
				& condensate charge $q_\star$
				\\
				spectral asymmetry $\mathcal{E}$
				& $\text{AdS}_2$ electric field $E$
			\end{tabular}
		\end{ruledtabular}
		\caption{
			\textbf{Dictionary of the Eliashberg-gravity map}. Summary of the correspondence between the degrees of freedom of the critical Eliashberg-SYK theory and those if (1+1)-dimensional holographic superconductor. Notice $\zeta$ and $\tau$ are related to $1/\epsilon$ and the abolute time by a Radon transformation.}
		\label{tab:dictionary}
	\end{table}
	Next we establish a holographic map based upon the effective action \eqref{eq:SYKscinit}. We first consider the case $T=\mu=0$.  It is convenient to only perform the Fourier transform $F\left(t,\varpi\right)$ with respect to the relative time and keep the absolute time $t$. From the Pauli principle for singlet, even-frequency pairing follows that $F\left(t,\varpi\right)$ does not depend on the sign of $\epsilon$. This allows us to introduce the scalar Bose field 
	\begin{equation}
	\tilde{\psi}\left(t,z\right)=c_{0}z^{\frac{\gamma-1}{2}}
	F\left(t,c/z\right),\label{eq:duality}
	\end{equation}
	with $z=c \left| \varpi \right|^{-1}$ and positive  coefficients $c$ and $c_{0}$.	Performing an inverse Radon transform $\tilde{\psi}\left(t,z\right)\rightarrow \psi\left(\tau, \zeta\right) $ -- a necessity pointed out in \cite{Maldacena:2016hyu,Das2018} -- that maps pairs of points $(t,z)$ to a point $(\tau,\zeta)$ on an AdS$_2$ geodesic, at low energy it then follows that the SYK theory takes the form 
	\begin{equation}
	S^{\left({\rm sc}\right)}/N=\int d \tau d\zeta \left(\frac{m^{2}}{\zeta^{2}}\left|\psi\right|^{2}+\left|\partial_{\tau}\psi\right|^{2}+\left|\partial_{\zeta}\psi \right|^{2}\right).
	\label{eq:adsform}
	\end{equation}
	This  is the the action of a holographic superconductor in AdS$_2$ with mass $m$ and with Euclidean signature:
	\begin{equation}
	ds^{2}=g_{ab}dx^{a}dx^{b}=\frac{1}{\zeta^{2}}\left(d\tau^{2}+d\zeta^{2}\right).
	\label{eq:T0metric}
	\end{equation}
	
	This derivation of a holographic map between SYK-Eliashberg and its AdS-dual, with the extra, radial direction $\zeta$ encoding the RG scale, is the central result of this paper.
	In the appendix  we give further details on the derivation of  Eq.\eqref{eq:adsform}, and  give explicit expressions for the  constants $c$ and $c_0$ as well as the mass $m$ in terms of the parameters of the Hamiltonian \eqref{eq:Hamiltonian}.
	The time derivative in Eq.\eqref{eq:adsform} originates from the first term in Eq.\eqref{eq:SYKscinit} and is due to the $\omega$-dependence of the particle-particle bubble $\Pi_{\rm n}(\omega,\varpi)$. The kinetic term in the radial direction of the holographic bulk is due to the singular pairing interaction $D_{\rm n}(\varpi)$.
	Both terms contribute to the AdS$_2$ mass $m$.
	The anomalous power-law dependence $\Pi_{\rm n}(0,\varpi)\sim |\varpi|^{\gamma-1}$   yields a positive contribution $m_{\left(1\right)}^{2}\propto \lambda_{p}^{-1}$.
	Thus, particle-particle fluctuations act against superconductivity and reducing  $\lambda_p$, further increases $m^2_{(1)}$. The finding  $m^2_{(1)}>0$ reflects the absence of a naive Cooper instability in critical non-Fermi liquids\cite{Balatsky1993,Sudbo1995}.
	However, the singular pairing interaction $D_{\rm n}(\varpi)\sim |\varpi|^{-\gamma}$  also leads to a negative contribution  $m_{\left(2\right)}^{2}<-\tfrac{1}{4}$ that favors pairing. It is, by itself, always more negative than the AdS$_2$
	Breitenlohner-Freedman (BF) bound $m_{{\rm BF}}^{2}=-\tfrac{1}{4}$ that signals the onset of an instability
	in AdS space \cite{Breitenlohner1982}.  
	The zero-temperature phase transition takes place when
	$m^{2}=m_{\left(1\right)}^{2}+m_{\left(2\right)}^{2}$ 
	equals $m_{{\rm BF}}^{2}$. This holographic condition is indeed identical to one  obtained from an analysis of the Eliashberg
	equation in ref.\cite{Hauck2020}.
	
	The holographic map was  established at zero temperature. At finite temperature one must periodically identify time in the SYK action, or equivalently work with discrete Matsubara frequencies. Holography, on the other hand, prescribes that we must change the background space-time metric to that of an Euclidean AdS$_2$ black hole
	\begin{align}
	ds^2 = \frac{1}{\zeta^2}\left((1-\zeta^2/\zeta_T^2)d\tau^2 + \frac{1}{(1-\zeta^2/\zeta_T^2)}d\zeta^2\right).
	\label{finite_T_metric}
	\end{align}
	Here $\zeta_T^{-1} = 2\pi T$ encodes the location of the black hole horizon.
	Because the normal state SYK physics in the IR is controlled by a conformal fixed point, the finite temperature theory at Euclidean time is mathematically directly related to the zero temperature one\cite{Kitaev2015,Sachdev2015}. The same is true in the AdS$_2$ theory: the black hole geometry is equivalent to the  $T=0$ background after a coordinate transformation\cite{Sachdev2019}. Tracing the transformation, the SYK model precisely translates to the finite $T$ version of the action \eqref{eq:adsform} with  metric Eq.\eqref{finite_T_metric}, see appendix.
	
	Finally we comment on the holographic map at finite chemical potential $\mu$. At low energies the normal state SYK model possesses an emergent U(1) symmetry\cite{Sachdev2015}. It allows to relate the normal state fermion propagator at finite $\mu$ to the one for $\mu=0$. For  $\Pi_{\rm n}(\omega,\varpi)$ this amounts to the shift $ \omega\to\omega-i4\pi q{\cal E}T$.
	On the other hand, the propagator of
	a charge $q^{*}$ scalar particle within AdS$_{2}$ that is exposed to a boundary electric field $E$, yields an analogous shift $\omega\rightarrow\omega-i2\pi q^{*}E T$.\cite{Faulkner:2009wj} Hence, the holographic boundary electric field  $E={\cal E}$ is given by the spectral asymmetry of the SYK model with effective charge $q^{*}=2q$ of the Cooper pair. 
	Within the gauge $A_{\zeta}=0$ this yields the vector potential $A_{\tau}=\tfrac{{ i}\cal E}{\zeta}\left(1-\zeta/\zeta_{T}\right)$.
	\begin{figure}[t]
		\centering
		\includegraphics[width=0.45\textwidth]{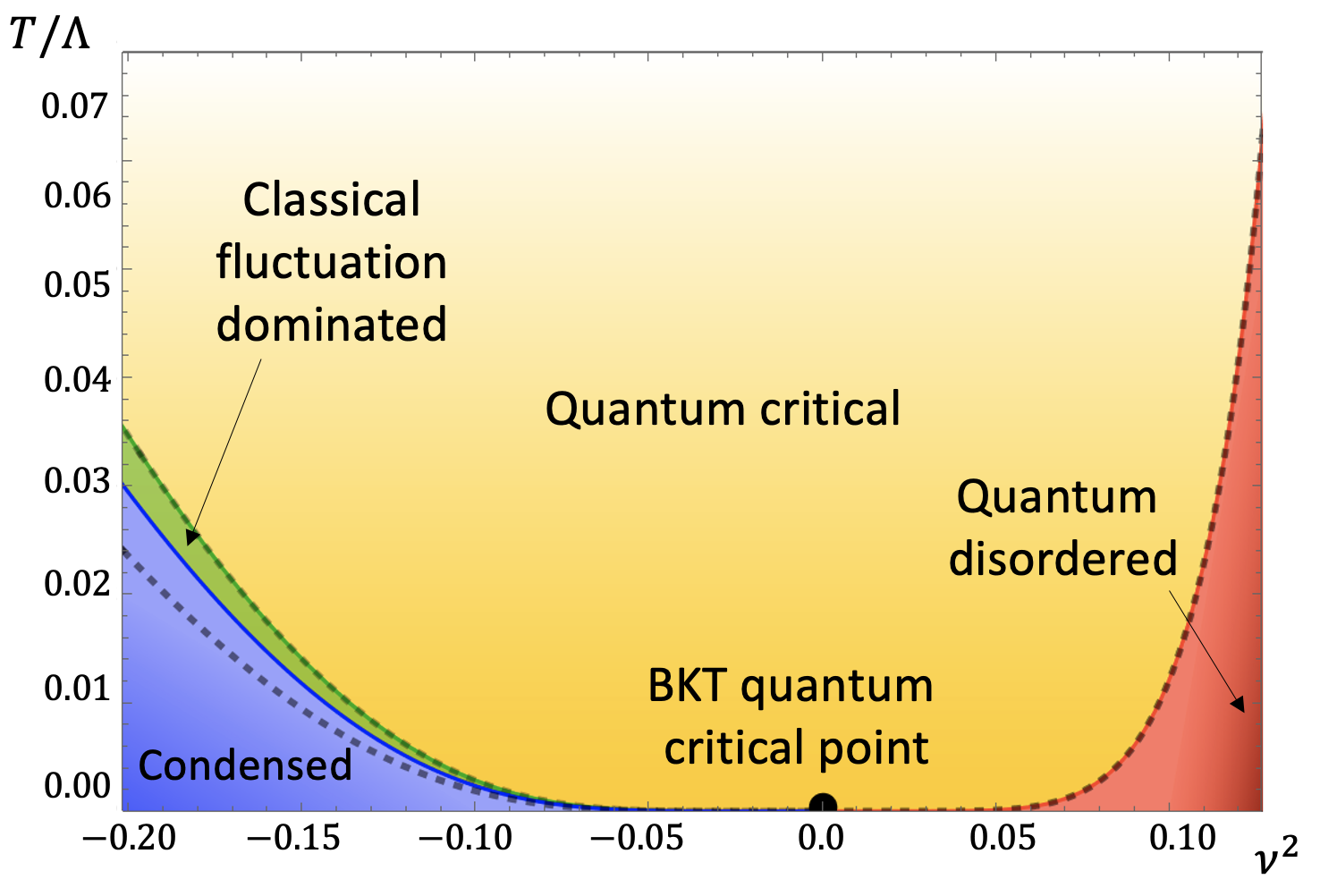}
		\caption{\textbf{Phase diagram of the SYK model}. Phase diagram as function of temperature $T$ and $\nu^2$ defined in Eq.\eqref{eq:nusquare} with $\gamma=0.9$. The susceptibility diverges at the transition temperature $T_{\rm c}$ (solid line) that vanishes in a BKT-like fashion at $\nu^2=0$. Below $T_{\rm c}$ we have superconducting order (blue). Between the dashed lines we find classical critical behavior with Curie-Weiss susceptibility and critical slowing down 
			$\chi\sim t_{\rm dyn}\sim (T-T_{\rm c})^{-1}$ (we only calculated the upper dashed curve, the lower one is an educated guess). As the susceptibility is finite at the BKT quantum critical point, quantum criticality is identified by the regime where $\partial_T\chi(T)\sim 1/(T \log^2T)$ (yellow region). in the quantum disordered region (red)  $ \partial_T\chi(T)$ saturates as $T\rightarrow 0$. The quantum critical and quantum disordered regimes are separated by a crossover.}
		\label{fig:phase_diagram}
	\end{figure}

	\subsection{Pairing response}
	
	Though we have equated the actions, this only implies that the dynamical equations of motions match. For a true equivalent mapping the response to an external source field must also correspond. We can do so by checking an explicit observable, the order parameter susceptibility in the normal phase. 
	We add an external pairing field $J_0(\tau)$ to the action:
	\begin{equation}
	S_{J}=-\int d\tau J_{0}\left(\tau\right)\sum_{i}c_{i\uparrow}\left(\tau\right)c_{i\downarrow}\left(\tau\right)+\text{h.c.}\, .
	\label{eq:source_SYK}
	\end{equation}
	$J_0(\tau)$ is conjugate to the dynamic order parameter $\mathcal{O}(\tau) = \lim_{\tau'\rightarrow \tau} \langle F(\tau,\tau')\rangle $ and physically realizable via coupling through a Josephson junction. Using our holographic map this yields in the gravitational formulation $S_J=-\int d^{2}x\sqrt{g}\left(J(x)\psi^{*}(x)+h.c. \right)$ where  at low frequencies $J\left(\tau,\zeta \right)=\frac{c}{c_{0}}\zeta^{\frac{1-\gamma}{2}}J_{0}\left(\tau\right)$. The corresponding Euler-Lagrange equation is
	\begin{equation}
	\partial_{\zeta}^{2}\psi+V\left(\omega,\zeta\right)\psi=\frac{J\left(\omega, \zeta\right)}{\zeta^2},\label{eq:stationary}
	\end{equation}
	with potential $V\left(\omega,\zeta\right)=-\frac{m^{2}}{\zeta^{2}}+\left(i\omega {+} 2q{\cal E}/\zeta\right)^{2}$.
	The $\zeta$-dependence on the r.h.s. of Eq.\eqref{eq:stationary} ensures that the source term, formally present in the AdS-bulk,  acts in essence only on the boundary, i.e. in the limit of small $\zeta$ dual to the UV, fully consistent with the holographic reasoning.
	
	The Euler-Lagrange Eq.\eqref{eq:stationary} should be identical to the stationary Eliashberg equation that follows from the original formulation of SYK model Eq.\eqref{eq:SYKscinit}. 
	Let us first consider a time independent source field $J_0$.  Expressed in terms of the anomalous self energy $\Phi(\varpi)=F\left(0,\epsilon\right)/\Pi_{\rm n}(0,\epsilon)$, the stationary equation for $F$ becomes
	\begin{equation}
	\Phi\left(\varpi\right)= \lambda_{p} \bar{\mu}_{\gamma}\int_{\varpi'}\frac{\Phi\left(\varpi'\right)}{\left|\varpi-\varpi'\right|^{\gamma} \left|\varpi'\right|^{1-\gamma}}+J_0,
	\label{eq:Eliashberg_lin}
	\end{equation}
	with $\bar{\mu}_{\gamma}$ given in the appendix. 
	Interestingly, this equation is identical to the linearized gap equation of the Eliashberg theory of numerous quantum-critical metals\cite{Abanov2001,Abanov2001b,Chubukov2005}. Thus, our  analysis goes beyond the specifics of the SYK model and is directly relevant to a much  broader class of physical systems. The power-law behavior in Eq.\eqref{eq:Eliashberg_lin} holds only within an upper (UV) and lower (IR) cut off $\Lambda$ and $T$, respectively. For $\left|\epsilon\right|\gg\Lambda$ the
	solution obviously becomes $\Phi\left(\epsilon\right)-J_{0}\propto\left|\epsilon\right|^{-\gamma}$, 
	while at $\left|\epsilon\right|\ll T$ the anomalous self energy should
	approach a constant. This fixes the UV and IR boundary conditions $\left.\frac{d}{d\epsilon}\epsilon^{\gamma}\Phi\left(\epsilon\right)\right|_{\Lambda}=\gamma\Lambda^{\gamma-1}J_{0}$
	and $\left.\frac{d}{d\epsilon}\Phi\left(\epsilon\right)\right|_{T}=0$.  Hence, we do  find that the source field acts through a UV boundary condition. As expected, the solution of the integral equation  \eqref{eq:Eliashberg_lin} and the much simpler Euler-Lagrange differential equation  \eqref{eq:stationary} at $\omega=0$ can be shown to be identical, provided we use the same boundary conditions.
	For small $\gamma$ it is in fact known that Eq.\eqref{eq:Eliashberg_lin} can be formulated as
	a differential equation\cite{Son1999,Chubukov2005,Hauck2020}. Our conclusion is, however not
	limited to small $\gamma$. 
	
	The $\omega=0$ solution of Eq.\eqref{eq:stationary} is 
	\begin{equation}
	\psi\left(0,\zeta\right)=A\zeta^{\frac{1}{2}-\nu}+B\zeta^{\frac{1}{2}+\nu},
	\end{equation}
	where $A$ and $B$ are determined by the boundary condition  translated for $\psi(\omega,\zeta)$.
	\begin{equation}
	\nu^2=\frac{1}{4}+m^{2}-\left(2q{\cal E}\right)^{2}
	\label{eq:nusquare}
	\end{equation}
	is a convenient measure of the mass and charge that vanishes at the BF bound.
	
	We can now easily determine the static susceptibility. Written with an eye on the holographic computations, it reads
	\begin{equation}
	\chi= \left. \frac{d {\cal O}  }{dJ_0} \right|_{J_{0}=0} =\chi_{\mathrm{{c}}}\frac{b_{+} + b_{-}\mathcal{G}(T)\Lambda^{-2\nu}}{a_{+}+a_{-}\mathcal{G}(T)\Lambda^{-2\nu}},\label{eq:uv-bc}
	\end{equation}
	where ${\cal G}\left(T\right)=c^{2\nu}\frac{B}{A}= \frac{2\nu-\gamma}{2\nu+\gamma}T^{2\nu}$  with $c$ from \eqref{eq:duality}
	contains the dependence on the IR behavior, while $a_{\pm}=b_{\pm}\left(1\pm2\nu/\gamma\right)^{2}$ and $b_{\pm}=1\mp2\nu/\gamma$  as well as $\chi_{\rm c}=2c_{g}^{2}\Lambda^{\gamma}/(\pi\gamma) $.
	
	The static susceptibility determines the phase diagram of the model shown in Fig.\ref{fig:phase_diagram}. The regime with superconducting order corresponds to   $\nu $ being imaginary. In that case there are solutions for the order parameter obeying the source-less $J_0=0$ boundary conditions. This occurs whenever the denominator in Eq.\eqref{eq:uv-bc} vanishes which determines the superconducting transition temperature:
	\begin{align}
	T_{{\rm c}} \sim\Lambda e^{-\frac{\pi}{\sqrt{-\nu^{2}}}}.
	\end{align}
	In Fig.\ref{fig:phase_diagram} the ordered state is shown in blue.  One recognizes a Berezinskii–Kosterlitz–Thouless (BKT) phase transition\cite{Kaplan2009} at a critical value of the pair-breaking parameter $\lambda_p $ that corresponds to $\nu=0$. Right above $T_c$ the susceptibility diverges according to the mean field Curie Weiss law $\chi(T)\sim(T-T_{\rm c})^{-1}$. This happens in the classical regime marked green in Fig.\ref{fig:phase_diagram}. 
	Note that at the BKT quantum criticial point the order parameter susceptibility is $\chi_{\rm c}$ and  stays finite. This  is in apparent contradiction with the behavior at quantum phase transitions, where commonly, the order parameter response to the external field is divergent. 
	However, the milder BKT quantum criticality is signaled by a diverging slope of the susceptibility as it approaches this finite value $\partial_T \chi(\nu=0)\approx  \tfrac{4\chi_{{\rm c}}}{\gamma T\log^{2}\left(\Lambda/T\right)}$  and $\partial_{\nu^2}\chi(T=0)\propto\chi_\text{c}/\nu$, fully consistent with results obtained from holography~\cite{Iqbal:2011aj,Jensen:2011af}.  The regime in the phase diagram where this is the behavior for $\partial_T \chi$ is indicated in yellow in Fig.\ref{fig:phase_diagram}, while a saturation of the slope happens below the crossover to the quantum-disordered regime (marked red).  
	
	\begin{figure}[t]
		\centering
		\includegraphics[width=0.35\textwidth]{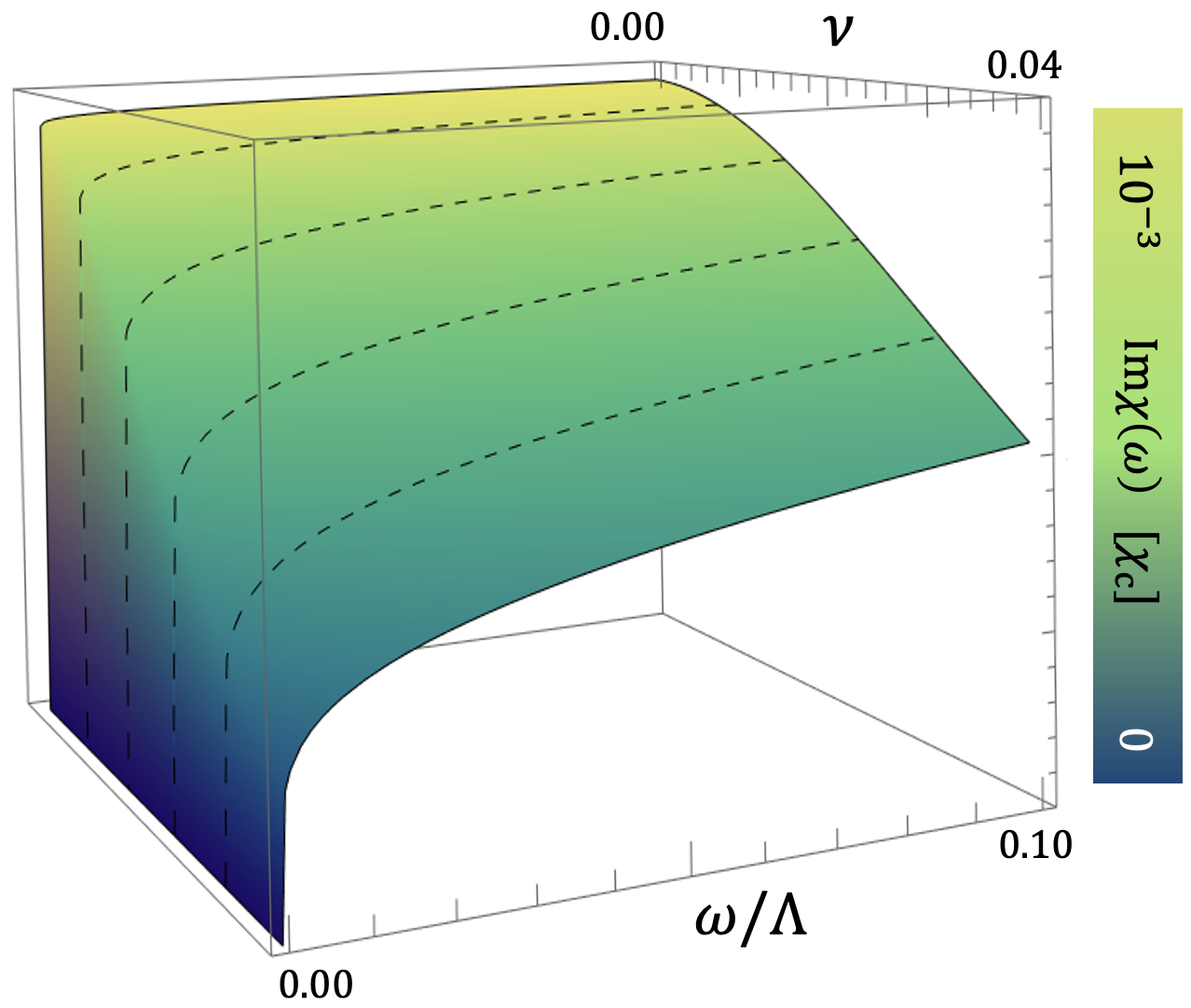}
		\caption{\textbf{Dynamical susceptibility away from criticality.} Imaginary part of the dynamic pairing susceptibility $\chi(\omega)$ for different values of the pair breaking, parametrized  in terms of $\nu$ with $\gamma=0.01$ and $T/\Lambda=10^{-4}$. Notice the broad continuum right at the BKT quantum critical point that only decays like $1/\log^2\omega$ at lowest frequencies. }
		\label{fig:dyn_suscept1}
	\end{figure}
	
\begin{figure}[t]
		\centering
		\includegraphics[width=0.40\textwidth]{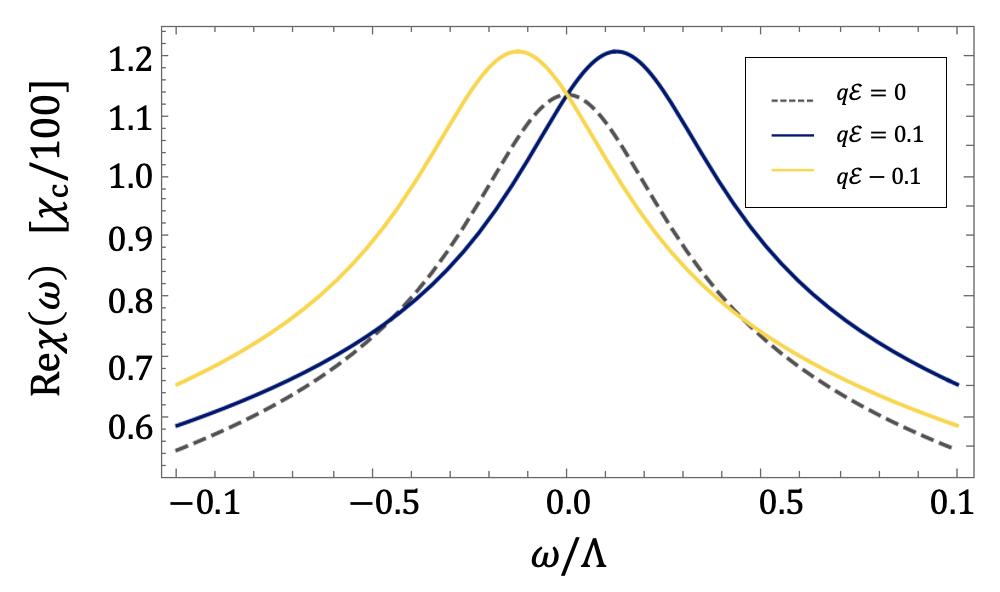}
		\caption{\textbf{Dynamical susceptibility away from charge neutrality.} Real part of the dynamic pairing susceptibility $\chi(\omega)$ for different values of boundary electric fields ${\cal E}$ corresponding to different carrier density with $\gamma=0.01, \nu=10^{-3}$, and $T/\Lambda=10^{-2}$. Notice spectral asymmetry away from $n=1/2$.}
		\label{fig:dyn_suscept2}
	\end{figure}

	Eq.\eqref{eq:uv-bc} is purposely written to be recognizable as 
	identical to the susceptibility  obtained from a holographic approach based on the crossover from a higher dimensional holographic theory AdS$_{d+2}$ at high energies to AdS$_{2}\times \mathbb{R}^{d}$ at low energies\cite{Iqbal:2011aj,Jensen:2011af}. Below we  discuss how the effect of the higher-dimensional space can be incorporated via appropriated double-trace deformations directly in the AdS$_2$ model Eq.\eqref{eq:adsform} \cite{Zaanen2015,Witten2001}.
	This insight allows us to use the power of holography and determine the dynamic pairing susceptibility. Using the approach of refs. \cite{Iqbal:2011aj,Jensen:2011af} we obtain $\chi(\omega)$ immediately from the static susceptibility of Eq.\eqref{eq:uv-bc} if we replace ${\cal G}\left(T\right)$ by 
	\begin{eqnarray}
	{\cal G}\left(T,\omega\right)=\frac{2\nu-\gamma}{2\nu+\gamma}T^{2\nu} \frac{\Gamma\left(u-\nu\right)\Gamma\left(v+\nu\right)}{\Gamma\left(u+\nu\right)\Gamma\left(v-\nu\right)}
	\label{eq:finite-T-ads2-greens}
	\end{eqnarray}
	with $u=\tfrac{1}{2}+i2q{\cal E}$ and $v=\tfrac{1}{2}-i\tfrac{\omega-4\pi Tq{\cal E}}{2\pi T}$. Here we use
	retarded functions, i.e. in-falling boundary conditions at the black hole horizon \cite{Faulkner:2009wj}.  In Fig.\ref{fig:dyn_suscept1} we show the resulting frequency dependence of the imaginary part of the pairing susceptibility at low $T$ as function of  $\omega$ for varying $\nu$. Most notable is the almost flat behavior right at the quantum critical point, where at lowest frequencies holds that ${\rm Im}\chi(\omega)\approx \tfrac{2 \pi}{\gamma} \log^{-2}\left(\Lambda/\omega\right)$. This  mirrors the temperature dependence discussed earlier.
	Near the classical transition at finite $T_{\rm c}$, we obtain critical slowing down behavior ${\rm Im}\chi(\omega)\sim\tfrac{\omega t_{\rm dyn}}{1+(\omega t_{\rm dyn})^2}$ with order-parameter relaxation time $t_{\rm dyn}\sim (T-T_{\rm c})^{-1}$. In Fig.\ref{fig:dyn_suscept2} we also show the impact of a deviation of the charge density from half filling, which induces a finite boundary electric field $\cal{E}$ and a spectral asymmetry of the two particle response.

	Finally we establish a more direct connection between the susceptibility of the SYK model and holography in AdS$_2$  without having to resort to crossover argument from higher-dimensional gravitational spaces. To this end
	we need to posit the correct holographic boundary conditions for AdS$_2$. In holography, boundary conditions at $\zeta=0$ crucially encode the precise theory one describes beyond the physics of interest within that theory. 
	For AdS$_2$ models this is a more detailed exercise, as the UV does not decouple as it does in higher dimensional theories \cite{Maldacena1999}, and further UV information is necessary to describe the theory. In brief: standard	boundary conditions correspond to a theory where $A = \lim_{\zeta\rightarrow 0}\zeta^{\nu-\frac{1}{2}}\psi(\zeta)\equiv J_0$ with $J_0$ the external source conjugate to the operator of interest, and $B=\lim_{\zeta\rightarrow 0}\frac{\zeta^{1-2\nu}}{2\nu}\partial_{\zeta}\zeta^{\nu-\frac{1}{2}}\psi|_{\zeta=0}=\mathcal{O}$ as the response; alternative quantization boundary conditions correspond to $B=J_0$, $A=\mathcal{O}$; a dual theory with so-called double trace deformations corresponds to mixed boundary conditions of the type $B = f A +J_0$, $B= \mathcal{O}$, where $f$ is an arbitrary coupling constant, or in alternative quantization\cite{Zaanen2015} $A= f B +J_0$, $A=\mathcal{O}$. When the UV does not decouple, however, there is a new, fourth possibility, where one includes the contribution of the leading UV operator. This corresponds to mixed boundary conditions $B = f A +J_0$ and a mixed response $B= g A +\mathcal{O}$ with again $g$ an arbitrary constant. The  susceptibility computed from AdS$_2$ is\cite{Faulkner:2009wj}  
	\begin{align}
	\chi_{\text{AdS}_2} =\left. \frac{d {\cal O}  }{dJ_0} \right|_{J_{0}=0}\simeq \frac{ \mathcal{O}}{J_0}= \frac{B- g A}{B-fA} = \frac{1-g\mathcal{G}}{1-f\mathcal{G}}.
	\label{eq:ads2-suscep}
	\end{align}
	Up to normalization, this equals the holographic result obtained via dimensional crossover\cite{Faulkner:2009wj} and the SYK susceptibility Eq.\eqref{eq:uv-bc}.

	\section{Summary}
	The low energy theory of the SYK Hamiltonian Eq.\eqref{eq:Hamiltonian} can thus be formulated in terms of the action 
	\begin{equation}
	S=S_{\rm gravity}+S_{\rm sc}+S_J.
	\label{eq:all_together}
	\end{equation}
	$S_{\rm gravity}$ is the much discussed AdS gravity in $1+1$ dimensions which eventually gives rise to the  Schwarzian theory, describing the SYK normal state \cite{Maldacena:2016hyu,Maldacena:2016upp}. It determines the metric of equations \eqref{eq:T0metric} and \eqref{finite_T_metric}. Hence the non-Fermi liquid normal state provides the gravitational background.  
	Superconducting degrees of freedom of the SYK model then behave like a charge $2q$ matter field on the AdS$_2$ background with:
	\begin{equation}
	S_{\rm sc}=N\int d^{2}x\sqrt{g}\left(D_{a}\psi^{*}D^{a}\psi+m^{2}\left|\psi\right|^2   \right),
	\label{eq:finaction}
	\end{equation}
	where $D_{a}=\partial_{a}{-}i2qA_{a}$. The finite temperature theory is equivalent to a change of coordinates that corresponds to a black hole with event horizon given by the Hawking temperature. Deviations of the density from half filling give rise to a  boundary electric field  ${\cal E}$.
	The competition between suppressed pairing due to the  absence of quasi-particles and the enhanced, singular pairing interaction is merely a balance of two contributions to the mass $m$ and leads to a quantum critical point of the BKT variety. The Euler-Lagrange equation of the gravitational theory maps directly onto the Eliashberg equation intensely studied in the context  of  quantum critical metals. External fields that couple to Cooper pairs act on the AdS$_2$ boundary. We have therefore  attained an explicit and complete AdS-field theory correspondence.
	
	Our derivation of holographic superconductivity from a microscopic Hamiltonian allows for a deeper and more concrete understanding of the holographic principle. Our  approach should be the natural starting point to study superconducting fluctuations beyond the Eliashberg limit and nonlinear effects, beyond the Gaussian level for systems as diverse as magnetic and nematic quantum critical points, or critical spin liquids. In the holographic language those effects correspond to quantum gravity corrections and gravitational back reactions, respectively. Moreover, as these systems are experimentally probed\cite{Chowdhury2019,Chubukov2020}, it provides a direct avenue to test holography in the lab.
	
	\section*{Acknowledgements}
	We are grateful to A. V. Chubukov for helpful discussions. 
	This research was supported in part by the Netherlands Organization for Scientific Research/Ministry of Science and Education (NWO/OCW), and the Deutsche Forschungsgemeinschaft (DFG, German Research Foundation) - TRR 288 -  422213477 Elasto-Q-Mat  (project A07).

	\section*{Appendix}
	\subsection*{A: SYK action of bilocal fields}
	In what follows we analyze the action of the SYK superconductor that follows from the Hamiltonian Eq.\eqref{eq:Hamiltonian}. 
	As is common practice in the theory of SYK like models we use bilocal fields, i.e. collective field variables that depend on two time variables. The effective action of the superconducting Yukawa-SYK problem is\cite{Esterlis2019,Hauck2020}:
	\begin{eqnarray}
	S & = & -N{\rm tr}\log\left(\hat{g}^{-1}-\hat{\Sigma}\right)+\frac{M}{2}{\rm tr}\log\left(d^{-1}-\Pi\right)\nonumber \\
	& - & N\left(G\otimes\Sigma+\widetilde{G}\otimes\widetilde{\Sigma}+F\otimes\Phi+F^{\dagger}\otimes\Phi^{\dagger}\right)+\frac{M}{2}D\otimes\Pi\nonumber \\
	& - & \frac{M}{2}\overline{g}^{2}\left((G\widetilde{G})\otimes D+\left(1-\alpha\right)\left(F^{\dagger}F\right)\otimes D\right),
	\end{eqnarray}
	with $M$ boson flavors and $N$ fermion flavors.  
	We use the notation
	$A\otimes B=\int d\tau_{1}d\tau_{2}A\left(\tau_{1},\tau_{2}\right)B\left(\tau_{2},\tau_{1}\right)$ as well as $F^{\dagger}\left(\tau_{1},\tau_{2}\right)=F\left(\tau_{2},\tau_{1}\right)^{*}$ and $\widetilde{G}\left(\tau_{1},\tau_{2}\right)=-G\left(\tau_{2},\tau_{1}\right)$.
	The fermionic self energy and propagator are a matrices in Nambu space:
	\begin{equation}
	\hat{\Sigma} =  \left(\begin{array}{cc}
	\Sigma & \Phi\\
	\Phi^{\dagger} & \widetilde{\Sigma}
	\end{array}\right) \,\,\, {\rm and}\,\,\, \hat{G}  =  \left(\begin{array}{cc}
	G & F\\
	F^{\dagger} & \widetilde{G}
	\end{array}\right).
	\end{equation} 
	The bare propagators are given as
	\begin{eqnarray}
	\hat{g}^{-1}\left(\tau,\tau'\right) & = & -\left(\partial_{\tau}\hat{\sigma}_{0}-\mu\hat{\sigma}_{z}\right)\delta\left(\tau-\tau'\right),\nonumber \\
	d^{-1}\left(\tau,\tau'\right) & = & -\left(\partial_{\tau}^{2}-\omega_{0}^{2}\right)\delta\left(\tau-\tau'\right),
	\end{eqnarray}
	where the $\hat{\sigma}_{i}$ are $2\times2$ Nambu-space matrices.
	For large $N$ and $M$, but fixed $N/M$ the exact solution of the
	single particle problem is given in terms of the stationary condition $\delta S=0$, which is given by the  Eliashberg equations:
	\begin{eqnarray}
	i\epsilon_{n}\left(1-Z\left(\epsilon_{n}\right)\right) & = & -\bar{g}^{2}\frac{M}{N}T\sum_{n^{\prime}}\frac{D\left(\epsilon_{n}-\epsilon_{n^{\prime}}\right)i\epsilon_{n'}Z\left(\epsilon_{n^{\prime}}\right)}{\left(\epsilon_{n'}Z\left(\epsilon_{n'}\right)\right)^{2}+\Phi\left(\epsilon_{n'}\right)^{2}},\nonumber \\
	\Phi\left(\epsilon_{n}\right) & = & \bar{g}^{2}\lambda_pT\sum_{n'}\frac{D\left(\epsilon_{n}-\epsilon_{n^{\prime}}\right)\Phi\left(\epsilon_{n^{\prime}}\right)}{\left(\epsilon_{n'}Z\left(\epsilon_{n'}\right)\right)^{2}+\Phi\left(\epsilon_{n'}\right)^{2}}.
	\label{eq:eliashberg}
	\end{eqnarray}
	Here, we considered $\mu=0$ and used $\hat{\Sigma}\left(\epsilon_{n}\right)=i\epsilon_{n}\left(1-Z\left(\epsilon_{n}\right)\right)\hat{\sigma}_{z}
	+\Phi\left(\epsilon_{n}\right)\hat{\sigma}_{x}$. This equation has to be supplemented by a corresponding expression for the boson self energy\cite{Esterlis2019}.
	
	In the normal state the expectation values of $\Phi$ and $F$ vanish. Hence, near the superconducting transition we can expand for small $\Phi$ and $F$. This yields
	\begin{eqnarray}
	S^{\left(sc\right)} & = & \Phi^{\dagger}\otimes G_{n}\otimes\Phi\otimes G_{n}-\frac{\bar{g}^{2}\lambda_{p}}{2}\left(F^{\dagger}F\right)\otimes D_{\rm n}\nonumber \\
	& - & F^{\dagger}\otimes\Phi-F\otimes\Phi^{\dagger}
	\label{eq:actionwithPhi}
	\end{eqnarray}
	where $G_{\rm n}$ and $D_{\rm n}$ are the finite temperature propagators of the
	normal state. If we now use that $G_{\rm n}$ and $D_{\rm n}$ only depend on the time difference, Fourier transform to frequency space, and integrate over the Gaussian fields $\Phi$ and $\Phi^\dagger$ we obtain Eq.\eqref{eq:SYKscinit} of the main text.
	
	The particle number $ n $ is related to the spectral asymmetry ${\cal E}$ though the generalized Luttinger theorem\cite{Georges2001,Wang2020}:
	\begin{equation}
	n=\frac{1}{2}-\frac{\theta}{\pi}-\frac{1-\gamma}{4}\frac{\sin\left(2\theta\right)}{\cos\left(\pi\gamma/2\right)},
	\end{equation}
	where $\tan\theta=\tan\left(\tfrac{\pi\left(1+\gamma\right)}{4}\right)\tanh\left(\pi q{\cal E}\right)$. 
	
	In the main text, we give in Eq.\eqref{norm-state-sols} the normal state propagators as function of time. Fourier transformation yields
	\begin{eqnarray}
	G_{\rm n}\left(\epsilon\right) & = & c_{g}\left(\tan\theta+i{\rm sign}\left(\epsilon\right)\right)\left|\epsilon\right|^{-\frac{1-\gamma}{2}}\nonumber \\
	D_{\rm n}\left(\epsilon\right) & = & c_{d}\left|\epsilon\right|^{-\gamma}
	\end{eqnarray}
	where the coefficients are
	\begin{eqnarray}
	c_{g}&=&2b_{f}\Gamma\left(\tfrac{1-\gamma}{2}\right)\cos\left(\tfrac{\pi}{4}(1+\gamma)\right)\nonumber \\
	c_{d}&=&2b_{b}\Gamma(\gamma) \cos(\pi \gamma/2),
	\end{eqnarray}
	and the exponent $\gamma$ is determined by
	\begin{equation}
	\frac{\left(1-\gamma\right)\tan\frac{\pi\left(1+\gamma\right)}{4}}{1+\tan^{2}\frac{\pi\left(1+\gamma\right)}{4}\tanh^{2}\left(\pi q{\cal E}\right)}=\frac{M}{N}\frac{\gamma\tan\frac{\pi\gamma}{2}}{1-\tanh^{2}\left(\pi q{\cal E}\right)}.
	\label{exponent_gamma}
	\end{equation}

	\subsection*{B: Derivation of the holographic map at $T=\mu=0$}
	First we summarize the derivation of  Eqs.\eqref{eq:duality} and \eqref{eq:adsform}.
	The analysis is straightforward for the first term in Eq.\eqref{eq:SYKscinit}. 
	We expand
	the zero-temperature expression of $\Pi^{-1}\left(\omega,\epsilon\right)$
	for small $\omega$ up to $\omega^{2}$: 
	\begin{eqnarray}
	\frac{1}{\Pi\left(\omega,\epsilon\right)} & \approx & \frac{1}{c_{g}^{2}}\left|\epsilon\right|^{1-\gamma}\left(1-\frac{1-\gamma}{8}\left(\frac{\omega}{\epsilon}\right)^{2}\right)
	\end{eqnarray}
	and Fourier transform from $\omega$ to Euclidean time $t$. One
	easily finds: 
	\begin{eqnarray}
	S^{\left({\rm sc},1\right)}/N & = & \int_{\omega,\epsilon}F^{\dagger}\left(\omega,\epsilon\right)\frac{1}{\Pi\left(\omega,\epsilon\right)}F\left(\omega,\epsilon\right)\nonumber \\
	& = & \int d z dt\left(\frac{m_{\left(1\right)}^{2}}{z^{2}}|\tilde{\psi}|^{2}-|\partial_{t}\tilde{\psi}|^{2}\right),
	\end{eqnarray}
	with $m_{\left(1\right)}^{2}=\frac{2\pi}{\lambda_{p} \bar{\mu}_{\gamma} b_{\gamma}}>0$. $b_\gamma$ will be defined below while \begin{equation}
	\bar{\mu}_{\gamma}=\mu_\gamma \Gamma\left(\gamma\right)\Gamma^2\left(\tfrac{1-\gamma}{2}\right)\left(2\cos\tfrac{\pi\gamma}{2}-\sin\left(\pi \gamma \right)\right). 
	\end{equation} 
	$\mu_\gamma$ is given in the main text.

	Next we analyze the second term
	\begin{equation}
	S_{{\rm SYK}}^{\left({\rm sc},2\right)}/N=-\frac{\bar{g}^{2}\lambda_{p}}{2}\int_{\omega,\epsilon,\epsilon'}F^{\dagger}\left(\omega,\epsilon\right)D\left(\epsilon-\epsilon'\right)F\left(\omega,\epsilon'\right)
	\end{equation}
	in  Eq.\eqref{eq:SYKscinit}. To proceed we perform a Mellin transform 
	\begin{equation}
	F\left(\omega,\varpi\right)=\int_{-\infty}^{\infty}\frac{dw}{\sqrt{2\pi}}f_{w}\left(\omega\right)
	\left|\varpi \right|^{i w-1+\gamma/2}
	\end{equation}
	and  the second term in  Eq.\eqref{eq:SYKscinit} becomes a set of uncoupled oscillators 
	\begin{equation}
	S_{{\rm SYK}}^{\left({\rm sc},2\right)}/N=\bar{g}^{2}\lambda_{p}c_d\int_{\omega,w}r_{w}f_{w}^{*}\left(\omega\right)f_{w}\left(\omega\right),
	\end{equation}
	where
	\begin{eqnarray}
	r_{w} &  = & \int_{-\infty}^{\infty}dx\frac{\left|x\right|^{iw}}{\left|1-x\right|^{\gamma}\left|x\right|^{1-\gamma/ { 2} }}\\
	& = & \frac{\pi^{2}/2}
	{\Gamma\left(\gamma\right)\cos\left(\frac{\pi\gamma}{2}\right)\left|\sinh\left(\frac{\pi}{4}\left(2w-i\gamma\right)\right)
		\Gamma\left(1+iw-\frac{\gamma}{2}\right)\right|^{2}}.\nonumber
	\end{eqnarray}
	Low energies corresponds to small $w$, where we can expand  
	\begin{equation}
	r_{w}=a_{\gamma}-b_{\gamma}w^{2}
	\end{equation}
	with positive coefficients
	\begin{eqnarray}
	a_{\gamma} & = & \frac{\pi^{2}}{2\Gamma\left(\gamma\right)\cos\left(\frac{\pi\gamma}{2}\right)\left(\sin\left(\frac{\pi}{4}\gamma\right)\Gamma\left(1-\frac{\gamma}{2}\right)\right)^{2}},\nonumber \\
	b_{\gamma} & = & \frac{\pi^{2}\left(\pi^{2}-4\psi^{\left(1\right)}\left(1-\frac{\gamma}{2}\right)\sin^{2}\left(\frac{\pi\gamma}{4}\right)\right)}{8\Gamma\left(\gamma\right)\Gamma^{2}\left(1-\frac{\gamma}{2}\right)\sin^{4}\left(\frac{\pi\gamma}{4}\right)\cos\left(\frac{\pi\gamma}{2}\right)}.
	\end{eqnarray}
	Inverting the Mellin transform finally yields in terms of the scalar field of Eq.\eqref{eq:duality}
	\begin{equation}
	-\frac{\bar{g}^{2}\lambda_{p}}{2}\int_{\omega,\epsilon,\epsilon'}F^{\dagger}DF=\int_{z,t}\left(\frac{m_{\left(2\right)}^{2}}{z^{2}}|\tilde{\psi}|^{2}+|\partial_{z}\tilde{\psi}|\right),
	\end{equation}
	with $m_{\left(2\right)}^{2}=-\frac{1}{4}-a_\gamma/b_\gamma$. Finally, the two coefficients in Eq.\eqref{eq:duality} are 
	\begin{eqnarray}
	c_{0}^{2}&=&\frac{1-\gamma}{8\pi c_{g}^{2}}c^{-\gamma},\nonumber \\
	c^{2}&=&\frac{\pi}{4}\frac{1-\gamma}
	{b_{\gamma}\overline{\mu}_{\gamma}\lambda_{p}}.
	\end{eqnarray}
	This completes the derivation of the action in Lorentzian de Sitter space with mass $m^2=m^2_{(1)}+m^2_{(2)}$.
	
	To transform the problem to AdS$_2$ we follow Ref.\onlinecite{Das2018} and use the Radon transform which takes the explicit form
	\begin{eqnarray}
	\tilde{\psi}\left(z,t\right)&=&\left({\cal R} \psi\right) (z,t)= 2z\int_{t-z}^{t+z}d\tau\int_{0}^{\infty}\frac{d\zeta}{\zeta}\psi\left(\zeta,\tau\right) \nonumber \\
	&\times& \delta\left(z^{2}-\left(\tau-t\right)^{2}-\zeta^{2}\right).
	\end{eqnarray}
	If one now uses that one can relate the Laplacian before and after the Radon transformation $\Box_{{\rm dS}_{2}}{\cal R}\psi=-{\cal R}\left(\Box_{{\rm AdS}_{2}}\psi\right)$ it follows that the Radon transform $\tilde{\eta}_\lambda={\cal R}\eta_\lambda$ of an eigenfunction $\eta_\lambda $ of the Laplacian in AdS$_2$ is also an eigenfunction of $\Box_{{\rm dS}_{2}}$. It is however not normalized. Addressing this issue through so-called leg factors\cite{Das2018} one obtains at low energy the action \eqref{eq:adsform} in AdS$_2$ with modified mass
	\begin{equation}
	\bar{m}{}^{2}=m_{{\rm BF}}^{2}+\frac{m^{2}-m_{{\rm BF}}^{2}}{\left[1-4G\left(m^{2}-m_{{\rm BF}}^{2}\right)\right]}.
	\end{equation}
	$G=\sum_{n=0}^{\infty}\frac{\left(-1\right)^{n}}{\left(2n+1\right)^{2}}\approx0.915966$ is Catalan's constant. Near the Breitenlohner-Freedman bound $m^2_{\rm BF}=\tfrac{1}{4}$ both masses are the same, which is the reason why we did not distinguish between $m$ and $\bar{m}$ in the main text.
	This completes	the holographic map.
	
	If we specify the density $n$, the ratio of boson and fermion flavors $M/N$, and the pair-breaking strength $\alpha$ the three crucial quantities ${\cal E}$, $\gamma$ and $\lambda_p$ are fixed and the gravitational theory is well defined. 
	Only one number - in our case $b_g$ - has to be fixed from a numerical solution of the SYK model. However it only determines the global  factor $c_0$ of the holographic action.

	\subsection*{C: Holographic map at finite temperatures}
	In the analysis at finite temperatures we start from the action for the SYK-superconductor Eq.\eqref{eq:actionwithPhi}
	without integrating out the conjugate field $ \Phi\left(\tau_{1},\tau_{2}\right)$. We have to keep in mind that the time integrations  go from $-\beta/2$ to $\beta/2$ where $\beta=1/T$.
	Similarly, $G_{{\rm n}}$ and $D_{{\rm n}}$ are the finite temperature
	propagators of the normal state. Using the invariance of the low-energy
	saddle point equations under re-parametrization $\tau\rightarrow\bar{\tau}=f\left(\tau\right)$,
	both functions can be obtained from the $T=0$ solutions $G_{{\rm n},0}\left(\tau\right)$ and
	$D_{{\rm n},0}\left(\tau\right)$ via 
	\begin{eqnarray}
	G_{\rm n}\left(\tau,\tau'\right) & = & f'\left(\tau\right)^{\frac{1+\gamma}{4}}G_{{\rm n},0}\left(f\left(\tau\right)-f\left(\tau'\right)\right)f'\left(\tau'\right)^{\frac{1+\gamma}{4}},\nonumber \\
	D_{\rm n}\left(\tau,\tau'\right) & = & f'\left(\tau\right)^{\frac{1-\gamma}{2}}D_{{\rm n},0}\left(f\left(\tau\right)-f\left(\tau'\right)\right)f'\left(\tau'\right)^{\frac{1-\gamma}{2}},\nonumber
	\end{eqnarray}
	with $f\left(\tau\right)=\frac{\beta}{\pi}\tan\left(\frac{\pi\tau}{\beta}\right)$.
	Hence, the finite temperature problem leads to an effective action,
	identical to the one we  analyzed at $T=0$, yet in terms of the
	field 
	\begin{eqnarray}
	\bar{F}\left(\bar{\tau}_{1},\bar{\tau}_{2}\right)&=&\left(1+\left(\pi T\right)^{2}\bar{\tau}_{1}^{2}\right)^{-\frac{1+\gamma}{4}}\left(1+\left(\pi T\right)^{2}\bar{\tau}_{2}^{2}\right)^{-\frac{1+\gamma}{4}} \nonumber \\
	&\times & F\left(f^{-1}\left(\bar{\tau}_{1}\right),f^{-1}\left(\bar{\tau}_{2}\right)\right)
	\end{eqnarray}
	Hence, we can immediately make the identification that is analogous
	to Eq.\eqref{eq:duality} only in terms of the new function $\bar{F}$
	instead of $F$ and with transformed variable $\bar{\tau}_{1,2}$ instead
	of $\tau_{1,2}$. The scalar field, expressed in terms of the anomalous propagator field, that replaces \eqref{eq:duality} at finite temperatures is then:
	\begin{eqnarray}
	\tilde{\psi}(\bar{t},\bar{z}) & = & c_{0}|\bar{\zeta}|^{\frac{1-\gamma}{2}}\int_{-\infty}^{\infty}ds\bar{F}\left(t_+,t_-\right)e^{ics/\bar{x}},
	\end{eqnarray}
	where $t_\pm=\bar{t}\pm\frac{s}{2}$. After performing an inverse Radon transformation from $\tilde{\psi}(\bar{t},\bar{z})$ to $\psi(\bar{\tau},\bar{\zeta})$,
	the holographic action can now be transformed from coordinates $(\bar{\tau},\bar{\zeta})$ to $\left(\tau,\zeta\right)$, where $0<\tau<1/T$, which agrees at the boundary with the original imaginary  variables of the SYK model\cite{Sachdev2019} 
	\begin{eqnarray}
	\bar{\tau} & = & 2\zeta_{T}\frac{\left(1-\zeta^{2}/\zeta_{T}^{2}\right)^{1/2}\sin\left(\tau/\zeta_{T}\right)}{1+\left(1-\zeta^{2}/\zeta_{T}^{2}\right)^{1/2}\cos\left(\tau/\zeta_{T}\right)},\nonumber \\
	\bar{\zeta} & = & \frac{2\zeta}{1+\left(1-\zeta^{2}/\zeta_{T}^{2}\right)^{1/2}\cos\left(\tau/\zeta_{T}\right)}.\label{eq:finiteTtransf}
	\end{eqnarray}
	where $\zeta_T=\tfrac{1}{2\pi T}$. In these coordinates the finite-$T$ action of the SYK superconductor
	becomes 
	\begin{eqnarray}
	S^{\left({\rm sc}\right)}/N&=&\int_{0}^{\zeta_{0}}d\zeta\int_{-\beta/2}^{\beta/2}d\tau\left(\frac{m^{2}}{\zeta^{2}}\psi^{*}\psi+\frac{\partial_{\tau}\psi^{*}\partial_{\tau}\psi}{1-\zeta^{2}/\zeta_{T}^{2}}\right. \nonumber \\
	&+& \left. \left(1-\zeta^{2}/\zeta_{T}^{2}\right)\partial_{\zeta}\psi^{*}\partial_{\zeta}\psi\right),
	\end{eqnarray}
	which is the correct finite-$T$ version of a holographic superconductor
	in AdS$_{2}$ with black hole horizon $\zeta_{T}$ and metric Eq.\eqref{finite_T_metric}.

\end{document}